\documentclass[prd,eqsecnum,apsi,twocolumn]{revtex4}

\usepackage{graphicx}
\usepackage{latexsym}
\usepackage{amsmath}
\usepackage{amssymb}

\newcommand{\be}{\begin{equation}}
\newcommand{\ee}{\end{equation}}
\newcommand{\bea}{\begin{eqnarray}}
\newcommand{\eea}{\end{eqnarray}}
\newcommand{\bfig}{\begin{figure}}
\newcommand{\efig}{\end{figure}}

\newcommand{\sech}{\hbox{sech}}

\begin{document}

\title{\Large \bf Neutrino masses, the cosmological constant and a stable 
universe in a Randall-Sundrum scenario}

\author{Paramita Dey${}^{}$ \footnote{E-mail: paramita@mri.ernet.in}, Biswarup
Mukhopadhyaya${}^{}$ \footnote{E-mail: biswarup@mri.ernet.in} }
\affiliation{Regional Center for Accelerator-based Particle Physics,\\
Harish-Chandra Research Institute, Chhatnag Road, Jhusi, Allahabad - 211 019,
India}

\author{Soumitra SenGupta${}^{}$ \footnote{E-mail: tpssg@iacs.res.in}}
\affiliation{ Department of Theoretical Physics and Center for Theoretical
Sciences,\\ Indian Association for the Cultivation of Science, Kolkata - 700
032, India}

\begin{abstract}
  The Randall-Sundrum model of warped geometry in a five-dimensional scenario,
  aimed at explaining the hierarchy between the Planck and electroweak scales,
  is intrinsically unstable in its minimal form due to negative tension of the
  visible brane. A proposed solution to the problem yields a negative
  cosmological constant in four dimensions. We show that this wrong-sign
  cosmological constant is restricted to small values, therefore requiring
  less cancellation from hitherto unknown physics, if bulk neutrinos are
  postulated to explain the observed neutrino mass pattern. Thus neutrino
  masses, a stable TeV-brane configuration and new physics in the context of
  the cosmological constant get rather suggestively connected by the same
  thread.
\end{abstract}

\maketitle

The Randall-Sundrum (RS) scenario provides an elegant explanation of the
hierarchy between the Planck ($M_P$) and electroweak (EW) energy scales
\cite{rs}. It postulates an extra spacelike compactified dimension, denoted by
$y~=~r_c\phi$, with two 3-branes at $\phi~=~0$ and $\pi$.  The latter contains
the physics of the standard model (SM) of elementary particles and is called
the `visible' brane.  The 5-dimensional `bulk' metric is given by
\be 
\label{metric}
ds^2 = e^{-2kr_c|\phi|} \eta_{\mu\nu} dx^\mu dx^\nu - r^2_c d\phi^2, 
\ee 
where $k$, of the order of the Planck mass, is related to the 5-d cosmological
constant, and $r_c$ is the brane-separation. The exponential `warp' factor
provides the aforementioned hierarchy of mass parameters, once projections on
the `visible' brane are taken, for $kr_c~\simeq~12$ \cite{rs}. Since this
allows all mass parameters in the 5-d theory, including $k$ and $1/r_c$ to lie
in the vicinity of the Planck scale, the solution can be deemed natural.

The entire set-up, however, is afflicted with a serious malady, namely,
negative tension for the visible brane \cite{genrs,gothers}. This makes the
brane configuration intrinsically unstable. A recent work \cite{genrs} has
suggested a solution to this problem by showing that one can take other
solutions to the warped geometry, where positive brane tension can be
achieved. This requires a negative bulk cosmological constant, as in the
original RS theory \cite{rs}, but also generates a non-zero cosmological
constant on the visible brane. 
While in the original Randall-Sundrum model the visible 3-brane is assumed to be flat with zero cosmological constant,
the possibility of a more generalized scenario with non-zero brane cosmological constant was explored in
a series of subsequent works \cite{earlygen}. Most of these works were focused in obtaining the effective gravitational theory on
an embedded hyper-surface in a bulk space-time and addressed the resulting cosmological issues. From a slightly different angle
 a generalized Randall-Sundrum model with non-zero brane cosmological constant was proposed by concentrating
on a proper resolution of the gauge hierarchy problem along 
with the possibility of having positive tension on the standard model TeV brane. It has been shown that
the effective (3+1)- dimensional Einstein's equation on the brane 
may be obtained with a non-zero brane cosmological constant (either positive or negative) where 
the TeV brane tension can be positive or negative, depending on the value of the brane cosmological constant. The  
solutions for the warp factors as well as the brane tensions were  
derived from the bulk equations
for both the de Sitter and anti-de Sitter cases. The modified warp factors which are functions of the 
induced cosmological constant on the brane tends to the Randall-Sundrum
exponential warp factor in the limit when the brane cosmological constant goes to  
zero. With such modified warp factor it is easy to obtain a Planck to TeV scale warping (to resolve the gauge hierarchy problem)
such that different values of the induced brane cosmological constants correspond to different values of brane
separation modulus $r_c$, for a given value of the parameter $k$ which is related to the bulk cosmological constant in the 
5-dimensional anti-de Sitter bulk space-time. 
This entire spectrum of solutions in the parameter space for both
de Sitter ( positive brane cosmological constant ) and anti-de Sitter (negative brane cosmological constant) regions have been
described in \cite{genrs} for this generalized Randall-Sundrum scenario. It has further been shown that for a wide range of
values of negative brane cosmological constant (i.e anti- de Sitter 3-brane), the brane tension for the TeV brane becomes
positive leading to a stable 3-brane configuration for a consistent embedding of the standard model on this brane. It may be 
recalled that in case of the original RS scenario the visible brane tension was necessarily negative and therefore was
intrinsically unstable.
In the anti-de Sitter region the magnitude of the `induced' 4-d
cosmological constant can be as high as upto $10^{-32}$ (rendered
dimensionless by scaling with $M^2_P$), which, with a concomitantly adjusted
$kr_c$, can lead to the requisite magnitude for the warp factor
\cite{genrs}. Since the cosmological constant is observationally restricted to
a very small value ($\simeq 10^{-124}$), this necessitates some new physics on
the 3-brane to cancel the induced negative cosmological constant. Obviously,
the smaller in magnitude is this induced brane cosmological constant, the less
is the demand on cancellation from new physics, thus making the scenario more
favourable. Moreover in this anti-de Sitter region, a large part of the parameter space
corresponds to a positive value for the visible brane tension and the resulting stability makes
it even more interesting in studying particle phenomenology. 
 
We show here that a constraint on the magnitude of the induced cosmological
constant does indeed follow, once this scenario is used to also generate
masses for neutrinos via massive bulk neutrinos
\cite{bulkferm,jd,gross,fermloc}. While the bulk neutrino mass(es) are
destined to be around the Planck scale, a sector of three light neutrinos is
generated, which essentially governs the neutrino mass patterns suggested by
the solar and atmospheric neutrino deficits \cite{sno,bando}. This mechanism,
including all its variants, has been discussed widely in the original RS
setting. Using the solution necessary for a positive brane tension, we find
that the constraints from neutrino masses favour relatively small magnitude of
the induced cosmological constant. Thus one can relate the apparently disjoint
issues of neutrino mass, the cosmological constant and a stable visible brane
in warped geometry. To state it from another angle, one strives to ensure
that the four-dimensional universe we live in is a stable configuration. In this
approach, the conduciveness of such a stable universe to structure formation has
a correspondence with tiny neutrino masses with a specific pattern. 

At this point we emphasize that
although the generalized RS scenario admits of a stable brane configuration by rendering the brane 
tension positive, the problem of modulus (brane separation parameter $r_c$) stabilization has to be addressed 
separately by introducing a bulk scalar field \cite{genrsstab} in the light of the  the modulus stabilization 
mechanism proposed by Goldberger-Wise and others \cite{gw}. In this mechanism 
it is shown that the bulk scalar field must have a non-zero vacuum expectation value (vev) on the branes in order to achieve
modulus stabilization. Since the bulk neutrinos considered here have no vev, they cannot act as stabilizing
fields for the brane separation modulus $r_c$.        

In general, allowing a non-vanishing but negative brane cosmological constant
(which has been shown to be a necessary condition for positive brane tension)
leads to a warp factor \cite{genrs,jd}
\be 
\label{warpf}
e^{-A(\phi)}          =     \omega         \cosh      \left(       \ln
  \frac{\omega}{1+\sqrt{1-\omega^2}} + kr_c\phi \right), 
\ee
where $\omega^2 ~=~ -\Omega^2/k^2$ is the absolute value of the dimensionless
quantity obtained out of the cosmological constant ($\Omega$) induced on the
brane. If one sets $e^{-A}~\simeq~10^{-16}$ to ensure the hierarchy between
the Planck and EW scales, then one can find two solutions for $kr_c\pi$ for
every $\omega^2$. No solution, however, exists for $\omega^2 > 10^{-32}$.  One
of the solutions, which yields the usual RS value of $kr_c\pi~\simeq 36.84$ in
the limit of near-vanishing $\omega$, always corresponds to a negative brane
tension, and hence is not relevant to our cause. The other solution, leading
to positive brane tension, gives increasing values of $kr_c\pi$ as $\omega^2$
decreases, leading to $kr_c\pi \simeq 250.07$ for $\omega^2 \rightarrow
10^{-124}$ (see Figures 1 and 2 in reference \cite{genrs}) On the whole, a
rather wide region in the $\omega^2 - kr_c\pi$ space is generally allowed,
which we set out to constrain from considerations of neutrino masses.

Out of the three neutrino mass eigenstates, the atmospheric and solar neutrino
deficits approximately suggest orders of mass-squared separation
\cite{sno,bando} as $\Delta m^2_{32} \simeq 10^{-20} - 10^{-21} ~{\rm GeV}^2$
and $\Delta m^2_{21} \simeq 10^{-22} - 10^{-23} ~{\rm GeV}^2$, which is
consistent with one massless physical state. In general, this allows a normal
hierarchy (NH) ($m_3 >> m_2 \simeq m_1$), an inverted hierarchy (IH)
($m_3\simeq m_2 >> m_1$), or degenerate neutrinos (DN) ($m_3 \simeq m_2 \simeq
m_1$). In addition, a bilarge mixing pattern has been strongly suggested by
the observed data \cite{neudata}.

The RS model allows for an explanation of neutrino masses without introducing
additional light neutrino states, by hypothesizing massive bulk neutrinos on
the order of the Planck mass \cite{bulkferm,gross}. One must choose a minimum
of two bulk neutrinos ($a = A,B$) for the cancellation of parity anomaly
\cite{parity}. The Kaluza-Klein (KK) decomposition of such a neutrino is
\be
\label{exp}
\Psi^{a}_{L,R} (x^\mu, y) = \sum_n
\psi^{a(n)}_{L,R}~(x^\mu)~\xi^{a(n)}_{L,R}~(y) 
\ee
where $\xi^{a(0)}_{L}~(y)$ can always be chosen to vanish identically (as we
require a right-handed zero mode on the visible brane). The function
$\xi^{a(0)}_{R}~(y)$ is obtained as
\be 
\label{soln}
\xi^{a(0)}_{R} (y) = \frac{N^{a(0)}}{\omega^2} \sech^2 \left( \ln
\frac{\omega}{1+\sqrt{1-\omega^2}} + ky \right) e^{-m_{a}y} 
\ee
where the normalisation constant is obtained from the condition $\int^\pi_0
d\phi r_c e^{-3A(\phi)} \xi^{a(0)*}_{R} \xi^{b(0)}_{R} = \delta^{ab}$. $m_A ,
m_B$ should each be at least $0.5k$, so that $\xi^{a(0)}_{R}$ is suppressed on
the visible brane, thus ensuring small neutrino masses. Non-vanishing
functions corresponding to both chiralities exist for the higher modes.
 
The neutrino masses are {\it prima facie} decided by the effective Yukawa
couplings $y^{(n)}_{a \ell}$ (for lepton flavour $\ell$) induced on the
brane. These are governed by the couplings $Y_{a\ell}$ involving the bulk
neutrino fields and the SM lepton and Higgs fields, and the value of
$\xi^{a(0)}_{R}$ at $\phi = \pi$. Once the compact dimension is integrated
out, the part of the action containing Yukawa interactions appears as
\be
\label{yukaction}
{\cal S}_{\rm Yukawa} = \int d^4x \frac{e^{-3kr_c\pi/2}}{\sqrt k}
Y_{a\ell} \bar{L_\ell} H \Psi_a (x,\pi),
\ee
and thus one has
\be
\label{yukfour}
y^{(n)}_{a \ell} = e^{3A(\pi)/2}~Y_{a\ell}~\xi^{a(n)}_{R}(\pi) / \sqrt{k}.
\ee
Including the KK mass tower $m^{(n)}$, the neutrino mass matrix takes the form
\be
\label{massm}
{\cal M}\ =\ \left(\! \begin{array}{ccc}
{\cal Y}^0_{3\times2} & {\cal Y}^1_{3\times2} & \cdots \\
0_{2\times2} & {\cal M}^1_{2\times2} & \cdots  \\
\vdots & \vdots & \ddots
\end{array}\!\right)\,,
\ee
where
\be
\label{massm1}
{\cal Y}^n_{3\times2}\ =\ v \left(\! \begin{array}{cc}
y^{(n)}_{Ae} & y^{(n)}_{Be} \\   
y^{(n)}_{A\mu} & y^{(n)}_{B\mu} \\
y^{(n)}_{A\tau} & y^{(n)}_{B\tau} \\
\end{array}\!\right)\,, ~ {\cal M}^n_{2\times2}~\ =\ \left(\! \begin{array}{cc}
m^{(n)}_{A} & 0 \\
0 & m^{(n)}_{B} \end{array}\!\right)\,.
\ee
The mass-squared matrix ${\cal M}{\cal M}^{\dagger}$ which ultimately decides
the mass eigenvalues for the light states, is
\be
\label{masssq}
{\cal M}{\cal M}^{\dagger}\ =\ \left(\! \begin{array}{ccc}
{\cal F}_{3\times 3} & {\cal G}^1_{3\times 2} & \cdots \\
{\cal G}^{1T}_{3\times 2} & ({\cal M}^1_{2\times2})^2 & \cdots\\
\vdots & \vdots & \ddots
\end{array}\!\right)\,,
\ee
with 
\be
\label{masssq1}
{\cal F}_{3\times 3}\ =\ v^2 \left(\! \begin{array}{ccc}
f_{ee} & f_{e\mu} & f_{e\tau} \\
f_{\mu e} & f_{e\mu\mu} & f_{\mu\tau} \\
f_{\tau e} & f_{\tau\mu} & f_{\tau\tau} \\
\end{array}\!\right)\,, 
\ee
where $v$ is the Higgs vacuum expectation value, $f_{\ell\ell'}= \sum_{n,a}
y^{(n)}_{a\ell}y^{(n)}_{a\ell'}$, and
\be
\label{masssq2}
{\cal G}^1_{3\times 2}\ =\ v \left(\! \begin{array}{cc}
y^{(1)}_{Ae}m^{(1)}_{A} & y^{(1)}_{Be}m^{(1)}_{B} \\
y^{(1)}_{A\mu}m^{(1)}_{A} & y^{(1)}_{B\mu}m^{(1)}_{B} \\
y^{(1)}_{A\tau}m^{(1)}_{A} & y^{(1)}_{B\tau}m^{(1)}_{B} \\
\end{array}\!\right)\,.
\ee
It is interesting to note that the eigenvalues of the light states, as
determined through the diagonalisation of ${\cal M}{\cal M}^{\dagger}$, are
decided only by the block ${\cal Y}^{0}_{3\times 2}$ in ${\cal M}$. One can
easily verify this when the ${\cal Y}$'s become c-numbers, i.e. there is just
one light neutrino which has a mass term in the top left-hand corner of the
mass matrix through coupling with a single right-handed state. In a way, this
is reminiscent of those cases in the type II seesaw mechanism \cite{type2ss}
where the left-handed Majorana mass prevails. However, this is just an
analogy, since the masses here are of Dirac type. It is straightforward to
check numerically that the same conclusion holds to a high degree of accuracy
for matrix-valued ${\cal Y}_{3\times 2}$, and the squared masses of the light
physical states are obtained just by diagonalising ${{\cal Y}^{0}_{3\times
2}}{{\cal Y}^{0}_{3\times 2}}^{\dagger}$. It is also noteworthy that the
procedure summarised above leads to the neutrino mass matrix derived in
\cite{gross}, in the limit of $\omega \rightarrow 0$, when the solution for
negative brane tension is taken (i.e. in the original version of RS).

One should also note that ${{\cal Y}^{0}_{3\times 2}}{{\cal Y}^{0}_{3\times
2}}^\dagger$ {\em always has one zero eigenvalue}. While this is consistent
with the NH and IH scenarios, degenerate neutrinos are disallowed in this
approach.

In our numerical study, we randomly vary all the $Y_{a\ell}$
$\ell~=~(e,\mu,\tau),~a~=~(A,B)$ in the range $0.01 - 5.0$ (this being a
`reasonable' range of dimensionless quantities), and $m_{A,B}$ in the range
$0.6k - k$, with $k~=~10^{19}$ GeV.  We demand that, in all cases,
$10^{-17}\le e^{-A(\pi)}\le 10^{-15}$. In order to fit the observed neutrino
pattern approximately, with a certain degree of tolerance, we have allowed
$\Delta m^2_{32} \simeq 10^{-20} - 10^{-21} ~{\rm GeV}^2$ and $\Delta m^2_{21}
\simeq 10^{-22} - 10^{-24} ~{\rm GeV}^2$. The neutrino mixing angles, answering
to the observed bilarge pattern, can be reproduced in specific combinations of
the $Y_{a\ell}$ included in the scan. Also, it is impossible to have all the
$Y_{a\ell}$ degenerate, irrespective of $m_A, m_B$, since that leads to two
zero mass eigenvalues. The other point to note is that the ratio $m_A/m_B$
cannot be widely hierarchical, as that, too, causes one of the columns in
${\cal Y}^0_{3\times 2}$ to be very small compared to the other due to the
exponential factor (see equations (\ref{soln}) and (\ref{yukfour})), once more
leading to two zero eigenvalues.
 \begin{figure}[htbp]
    \begin{center}
      {\resizebox{4.7cm}{!}{\includegraphics{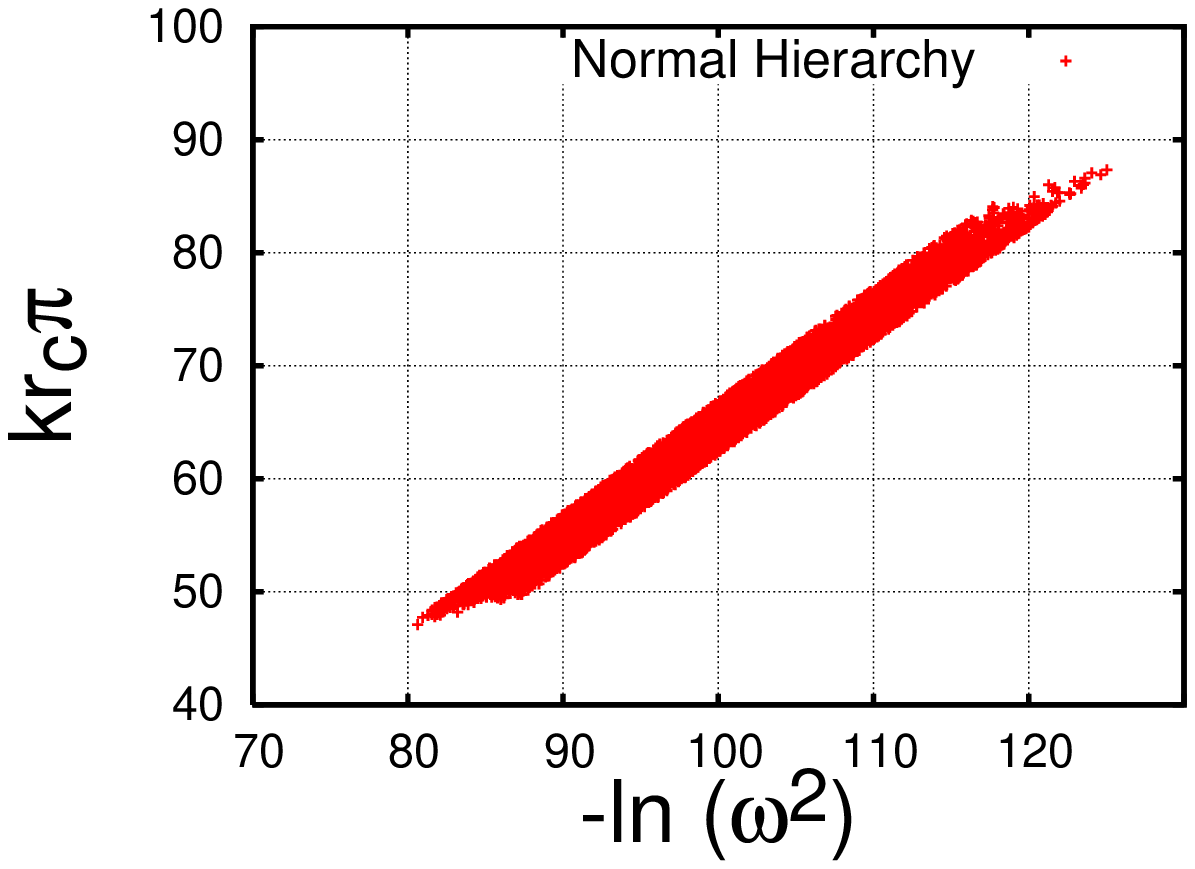}}}
       \hspace*{-1.40cm}
      {\resizebox{4.7cm}{!}{\includegraphics{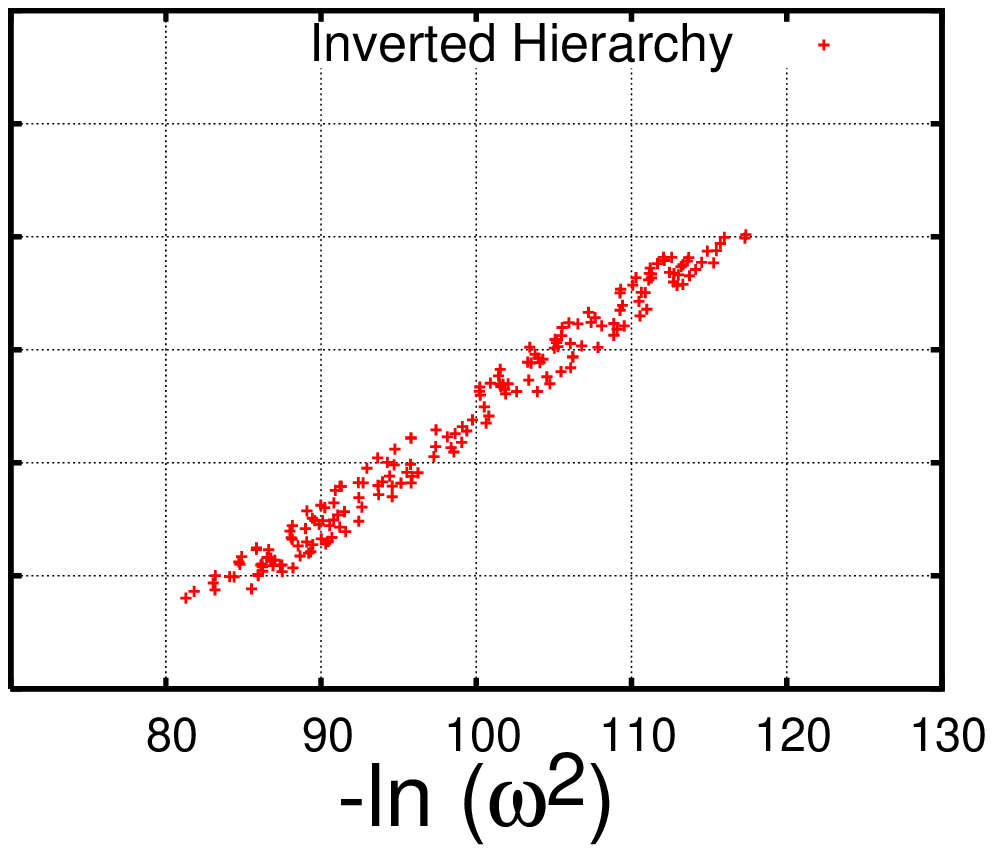}}}
     \caption{Areas in the $\omega-kr_c\pi$ plane for positive tension of
     visible brane allowed by the NH (left) and IH (right) scenarios of
     neutrinos.}
\label{fig1}
\end{center}
\end{figure}

The results of our numerical analysis, yielding constraints on the parameter
space, are summarised in Figures 1 - 3. Figure 1 shows the areas in the
$\omega^2 - kr_c\pi$ plane allowed by the NH and IH scenarios in turn.  Most
remarkably, out of the erstwhile allowed range of $\omega^2$ reaching upto
$10^{-32}$, only the region corresponding to $\omega^2\lesssim 10^{-80}$ is
allowed in both the NH and IH cases.

While $\omega^2$ can in principle be arbitrarily small, we find, with our
range of Yukawa couplings chosen, increasingly fewer solutions as one goes to
$\lesssim 10^{-120}$. In any case, for solutions corresponding to positive
brane tension, values of $\omega^2$ smaller than this leads to a rather wide
hierarchy between $k$ and $1/r_c$, as can be seen from Figure 1. {\em The
absence of solutions for very low values of $\omega^2$ highlights a very
important role of neutrinos here: for the positive tension solution, one
cannot consistently generate neutrino masses with arbitrarily small magnitude
of the induced cosmological constant. This in turn rules out a large hierarchy
between $k$ and $1/r_c$. Phenomenology of the neutrino sector thus renders
such a scenario `natural'.}

 \begin{figure}[htbp]
    \begin{center}
      {\resizebox{5cm}{!}{\includegraphics{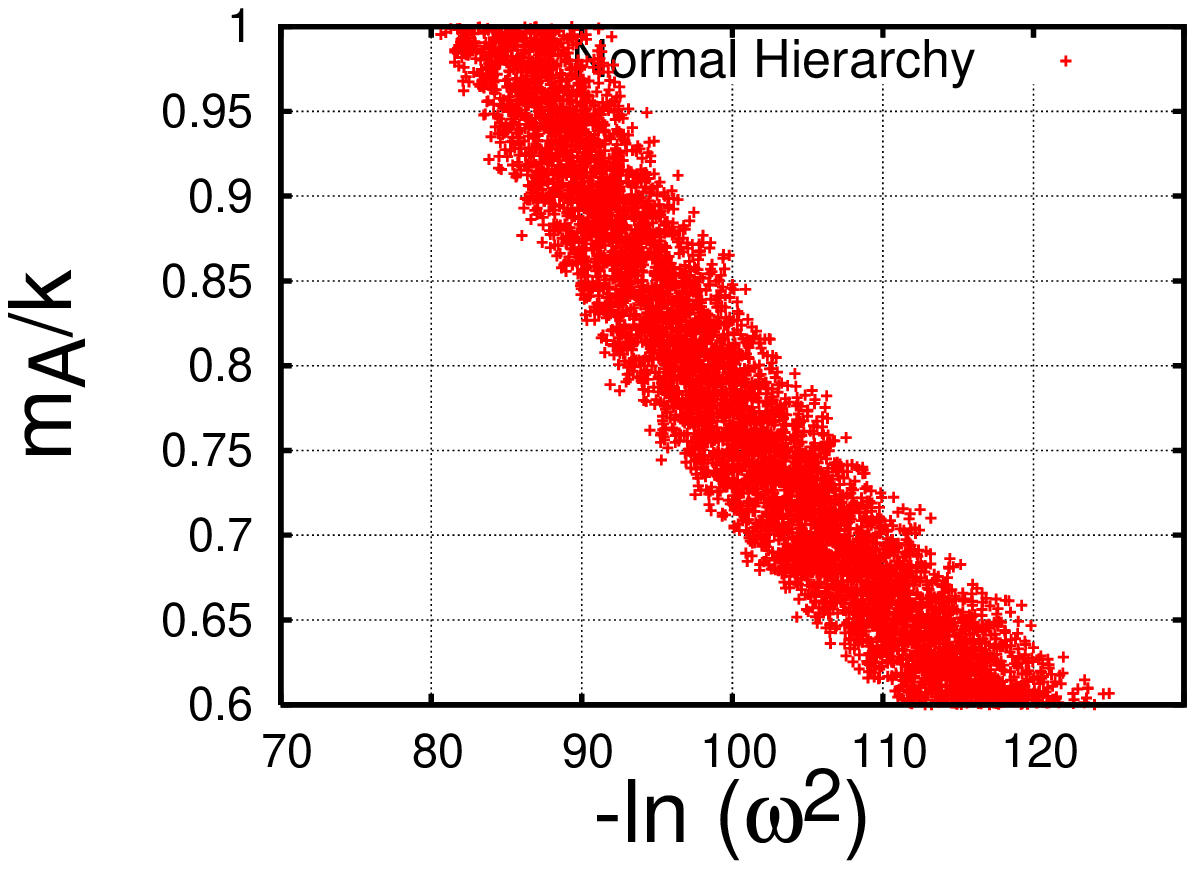}}}
      \hspace*{-1.6cm}
      {\resizebox{5cm}{!}{\includegraphics{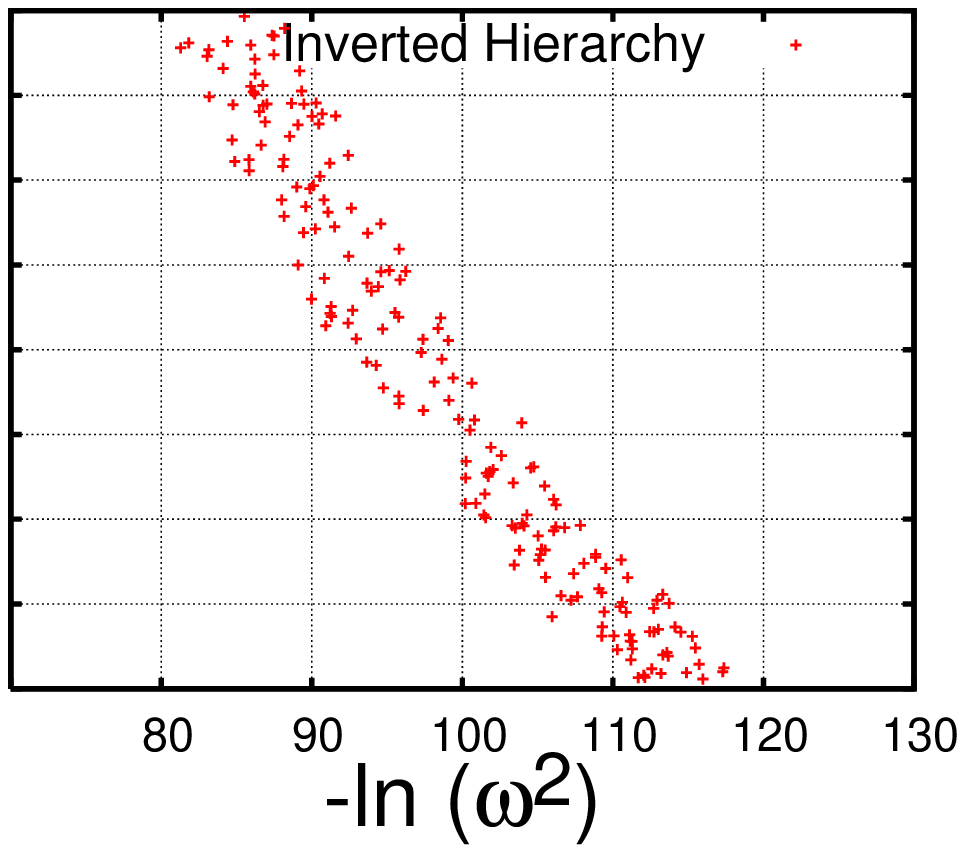}}}
      \vspace*{-0.3cm}\begin{center}{\bf (a)}\end{center}
      {\resizebox{5cm}{!}{\includegraphics{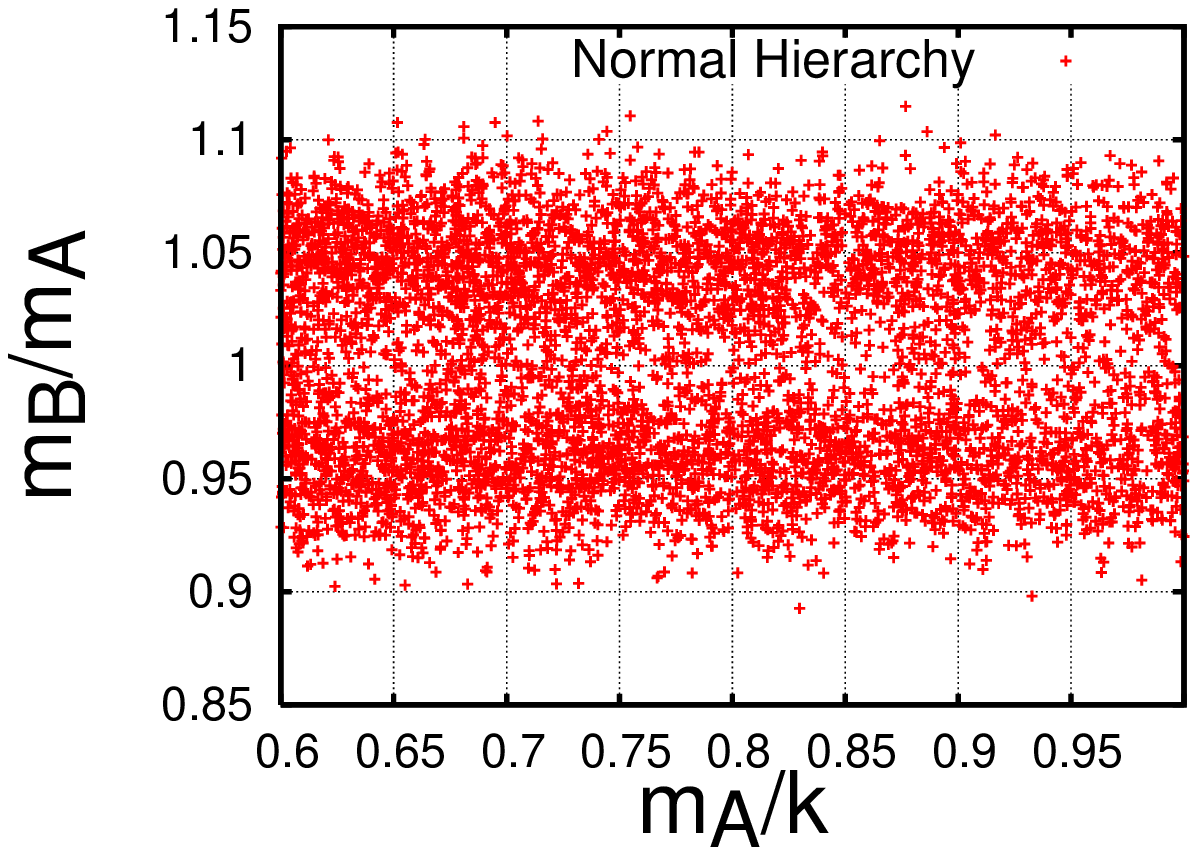}}}
      \hspace*{-1.6cm}
      {\resizebox{5cm}{!}{\includegraphics{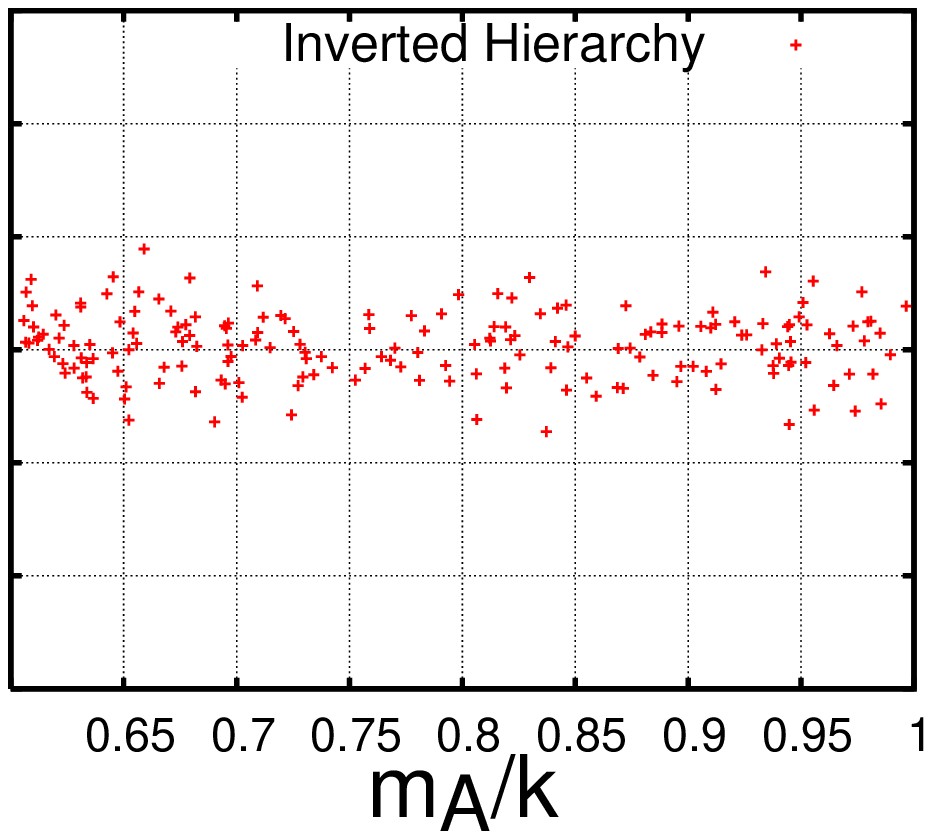}}}
      \vspace*{-0.3cm}\begin{center}{\bf (b)}\end{center}
      \label{fig2}
     \caption{The ranges of $m_{A,B}$ allowed by the NH (left panel) and IH
     (right panel) scenarios {\em{vis-a-vis}} -ln$(\omega^2)$.}
\end{center}
\end{figure}

Figure 2 shows the ranges of $m_{A,B}$ allowed by our solutions which include
the whole range over which we have varied the masses.  However, as is evident
from Figure 2(b), the ratio of the two bulk masses is constrained to be in the
approximate range $0.9 - 1.1$.  The plots are symmetric about $m_A = m_B$,
since the labeling of the two bulk neutrinos is arbitrary. This confirms our
earlier statement that a wide hierarchy between the two bulk masses is
disallowed. Also, there is a depletion of available points around $m_A = m_B$,
as one then needs the $Y_{a\ell}$'s to be sufficiently apart.
 \begin{figure}[htbp]
     \begin{center}
      {\resizebox{5cm}{!}{\includegraphics{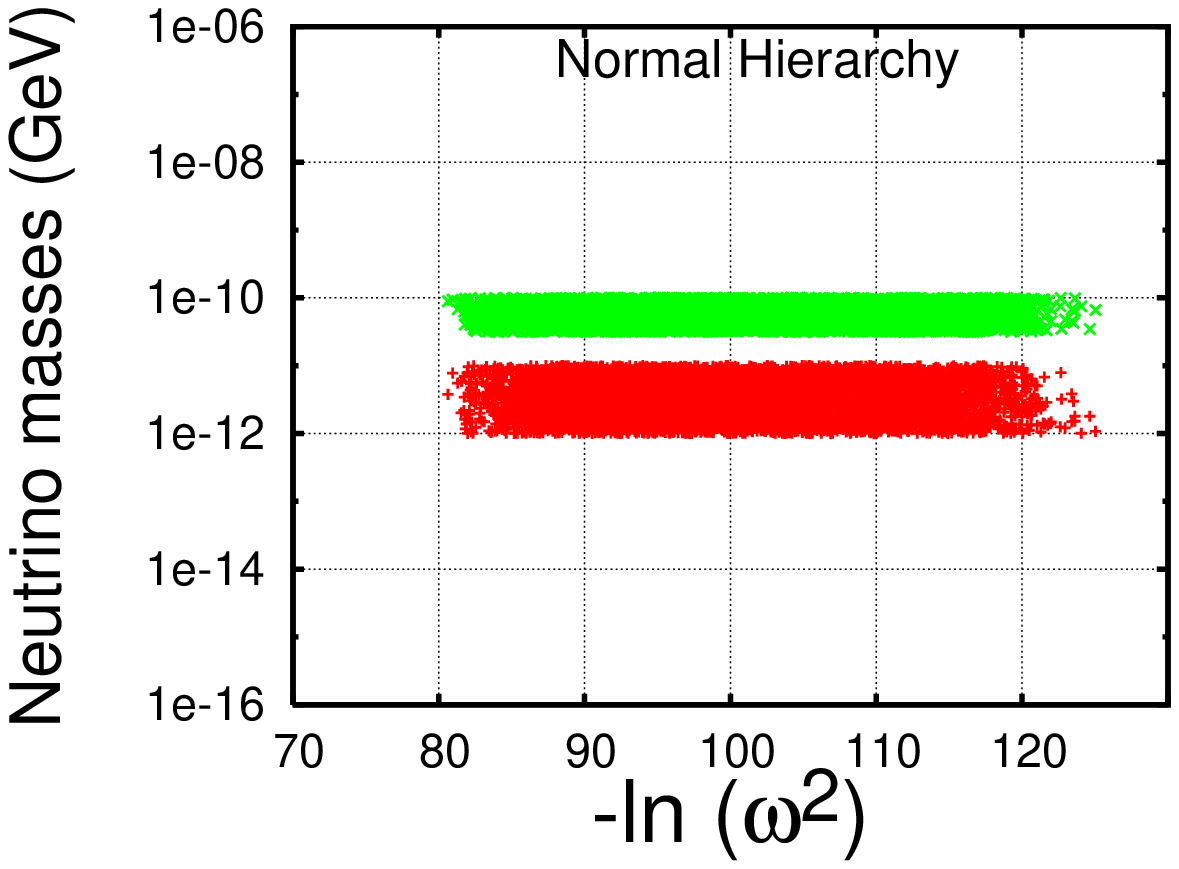}}}
      \hspace*{-1.68cm}
      {\resizebox{5cm}{!}{\includegraphics{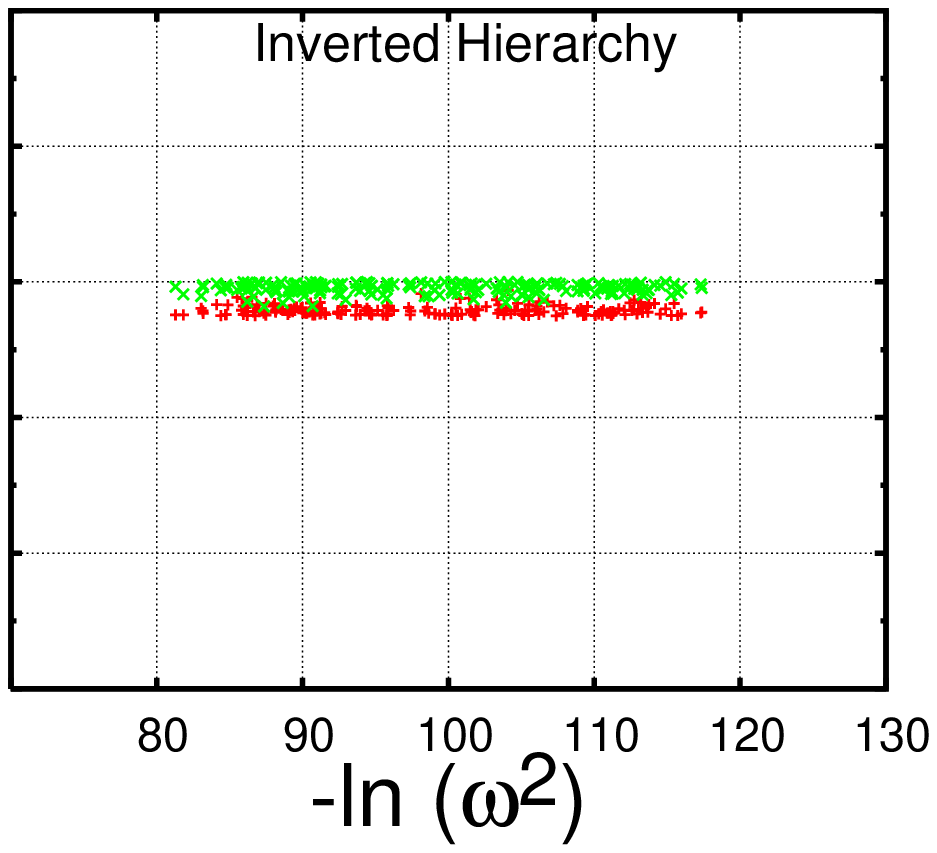}}}
     \caption{Allowed ranges of neutrino masses against -ln$(\omega^2)$, for
     NH (left) and IH (right). One mass eigenvalue is always zero in each
     case.}
\label{fig3}
\end{center}
\end{figure}

Finally, Figure 3 displays the allowed bands of neutrino masses for NH and IH,
plotted against $\omega^2$. One mass eigenvalue is zero in each case, due to
the nature of the induced mass matrix. It is clear again from these plots that
one has the requisite mass patterns for $10^{-120}\lesssim \omega^2 \lesssim
10^{-80}$ only, so long as the solutions for positive brane tension are taken.
It is also obvious from this figure, as much as from the previous ones, that
one gets much denser scatter plots for NH as compared to IH.  This is because
one of the eigenvalues is zero from the very structure of the induced mass
matrix, and one requires a finer split between the two non-vanishing
eigenvalues for IH than for NH. Thus the NH scenario is somewhat more favoured
than IH, as evinced from our scatter plots.

It is also possible to turn the above conclusions around in the following way.
While the requirement of a positive tension brane can lead to values of
$\omega^2$ as high as $10^{-32}$, observation demands the value to be close to
$10^{-124}$. One may consider it desirable to keep the generated value of
$\omega^2$ at the latter value, unless a separate source to cancel the large
induced value can be invoked.  If that indeed be the case, and if a mechanism
of neutrino mass generation has to be envisioned in terms of bulk sterile
neutrinos in a warped geometry, then the bulk masses of these states have to
lie in the range of $0.6 k$, and a hierarchy of about 80 between $k$ and
$1/r_c$ is likely.

In summary, we find that the requirement of fitting the observed neutrino mass
pattern in terms of bulk neutrinos imposes stringent constraints on the value
of the cosmological constant on the visible brane, if one has to obtain a
positive brane tension in an RS theory.  More specifically, the magnitude of
the induced cosmological constant (inheriting a negative sign), is restricted
to be less than $10^{-80}$ in dimensionless units, thus bringing down its
upper limit by 48 orders compared to what is allowed without any reference to
neutrinos.  The issue of the cosmological constant is admittedly open in most
theories beyond the standard model of elementary particles, and some hitherto
unknown physics is expected to play a role in its small positive value. In the
RS context, our study shows that the necessity of explaining neutrino masses
minimises the required level of fine tuning to reproduce the observed value of
the cosmological constant, presumably through some yet undiscovered feature of
the 3-brane that constitutes our visible universe. Alternatively, one obtains
constraints on the bulk neutrino masses, and also on the hierarchy between $k$
and $1/r_c$, if there is no other mechanism for explaining the observed value
of the cosmological constant. Thus neutrino masses get tantalisingly linked
with the issue of the cosmological constant.  And remarkably, a positive
tension brane, or a stable four-dimensional universe within the RS framework,
emerges triumphant while such a link is established.

\acknowledgements{The work of PD and BM was partially supported by funding
available from the Department of Atomic Energy, Government of India, for the
Regional Center for Accelerator-based Particle Physics, Harish-Chandra
Research Institute. PD and BM also acknowledge the hospitality of Indian
Association for the Cultivation of Science, Kolkata, while the work was in
progress.}


\end{document}